\documentclass[conference]{IEEEtran}
\IEEEoverridecommandlockouts
\usepackage{graphicx}
\usepackage{amsmath,amssymb,amsfonts}
\usepackage{mathtools}
\usepackage{bbm}
\usepackage{cite}
\usepackage{bm}
\usepackage{siunitx}
\usepackage{algorithm}
\usepackage{algorithmic}
\usepackage{stfloats}
\usepackage{subfigure}
\usepackage{textcomp}
\usepackage{xcolor}

\newtheorem{lemma}{\textbf{Lemma}}

\begin{document}
\bstctlcite{IEEEexample:BSTcontrol}
\title{Event-triggered Online Proactive Network Association to Mobile Edge Computing for IoT}
\author{\IEEEauthorblockN{Jian Zhang\IEEEauthorrefmark{1}, Qimei Cui\IEEEauthorrefmark{1}, Xuefei Zhang\IEEEauthorrefmark{1}, Kwang-Cheng Chen\IEEEauthorrefmark{2}, and Ping Zhang\IEEEauthorrefmark{3}}
\IEEEauthorblockA{\IEEEauthorrefmark{1}National Engineering Lab for Mobile Network Technologies, \\
\IEEEauthorrefmark{3}State Key Laboratory of Networking and Switching Technology,
\\Beijing University of Posts and Telecommunications, Beijing, 100876, China \\
\IEEEauthorrefmark{2}Department of Electrical Engineering, University of South Florida, Tampa, FL, 33620, USA
\\Email:\{bupt\_zhangjian, cuiqimei, zhangxuefei\}@bupt.edu.cn; kwangcheng@usf.edu; pzhang@bupt.edu.cn}}

\maketitle
\begin{abstract} 
Ultra-low latency communication for mobile machines emerges as a critical technology in Internet of Things (IoT). Proactive network association has been suggested to support ultra-low latency communication with the assistance of mobile edge computing. To resolve system dynamics and uncertainty, in this paper, an online proactive network association is proposed to minimize average task delay while considering time-average energy consumption constraints. Under distributed computing and networking environments, we formulate an event-triggered proactive network association model by semi-Markov task states and independent identically distributed (i.i.d.) random events. Then we facilitate the mobility-aware anticipatory network association to predictively consider handover effects caused by the mobility. Based on the Markov decision processes (MDP) and Lyapunov optimization, the two-stage online proactive network association (TOPNA) decision algorithm is proposed without the knowledge nor distribution of random events. Simulation results exhibit the effectiveness of the proposed algorithm.   
\end{abstract}
\section{Introduction}
With the rapid development of Internet of Things (IoT) applications, such as intelligent transportation, smart city and autonomous vehicles, etc. 
Ultra-low latency communication for mobile machines is critical to achieve system reliability \cite{ckc2018ultra}. 
For example, real-time vision analysis of surrounding road environment perceived by automatic vehicles to avoid potential traffic accidents.
In recent years, mobile edge computing (MEC) has emerged to enable mobile applications due to its potential of achieving low delay and saving energy consumption through pushing computing capabilities to network edges \cite{hu2015mobile}. 
Proactive network association therefore plays an important role to support ultra-low latency or delay-sensitive services under the assistance of MEC or fog network \cite{hung2018delay}.

Different from traditional cloud computing with stable dedicated wired connection, mobile machines encounter highly dynamic and stochastic network environment, such as time-varying channel quality and unpredictable available server's computing resource. However, most existing researches on network association ignore the stochasticity and unpredictability nature of mobile machines in such a network architecture. For example, in \cite{wang2016joint,ye2013user}, combinatorial optimization problem are formulated to myopically maximize transmission rates, which can be solved by distributed decomposition methods. In \cite{trestian2012game}, a comprehensive survey of game theoretic approaches on network association is presented to choose the maximization of myopic utility between different strategies.

Some recent works have made some efforts to maximize long-term reward of network association in order to deal with the stochasticity of network environments. In \cite{wang2016network,ko2018spatial}, Markov decision processes (MDP) are proposed to maximize the total amount of network association reward in the long run, which can be solved by linear programming.  
In \cite{li2016joint,lyu2017optimal}, Lyapunov optimization technologies are leveraged to optimize time-averaged energy consumption and network throughput accordingly by transforming stochastic optimization into deterministic optimization over timeslots. 

However, the aforementioned studies treat decision-making of network association based on the discrete timestep or slotted structure with fixed and equal decision intervals, which violates the practical scenarios of operating mobile machines that the network association depending on the task completion duration \cite{heemels2012introduction}. Moreover, those investigations assume that the stochasticity of network environment satisfies the Markov dynamics or independent and identically distributed (i.i.d.) properties separately, which neglects the existence of both Markov property states (i.e., mobility trajectory and network association) and i.i.d. environment nature (i.e., channel quality and computing resource) in the network association problem, to over-simplify the low-latency operation and system reliability.

Facilitated by machine learning and autonomous operations by edge devices or smart machines/agents, proactive network association is critical in distributed computing scenarios, such as automatic driving and emergency rescue, by eliminating reliance on centralized decision-making entity \cite{li2017distributed}. In this paper, we develop the online proactive network association decision based on event-triggered distributed computing environment, to minimize the average task delay subjected to time-averaged energy consumption for mobile machines in IoT, particularly those mobile machines of artificial intelligence. The main contributions of this work are as follows:
\begin{itemize}
	\item Different from the tradition network association model based on slotted structure, we develop an event-triggered proactive network association decision model under distributed computing environment by taking semi-Markov task states and i.i.d. random events into account. 
	\item We extend the above model to the mobility-aware anticipatory network association decision problem by considering the further impacts on handover triggered by mobility. 
	\item Based on the MDP and Lyapunov optimization theory, we propose the two-stage online proactive network association (TOPNA) decision algorithm. Simulation results show the effectiveness of our proposed algorithm.  
\end{itemize}
\begin{figure}[t]
	\centering
	\includegraphics[width=0.45\textwidth]{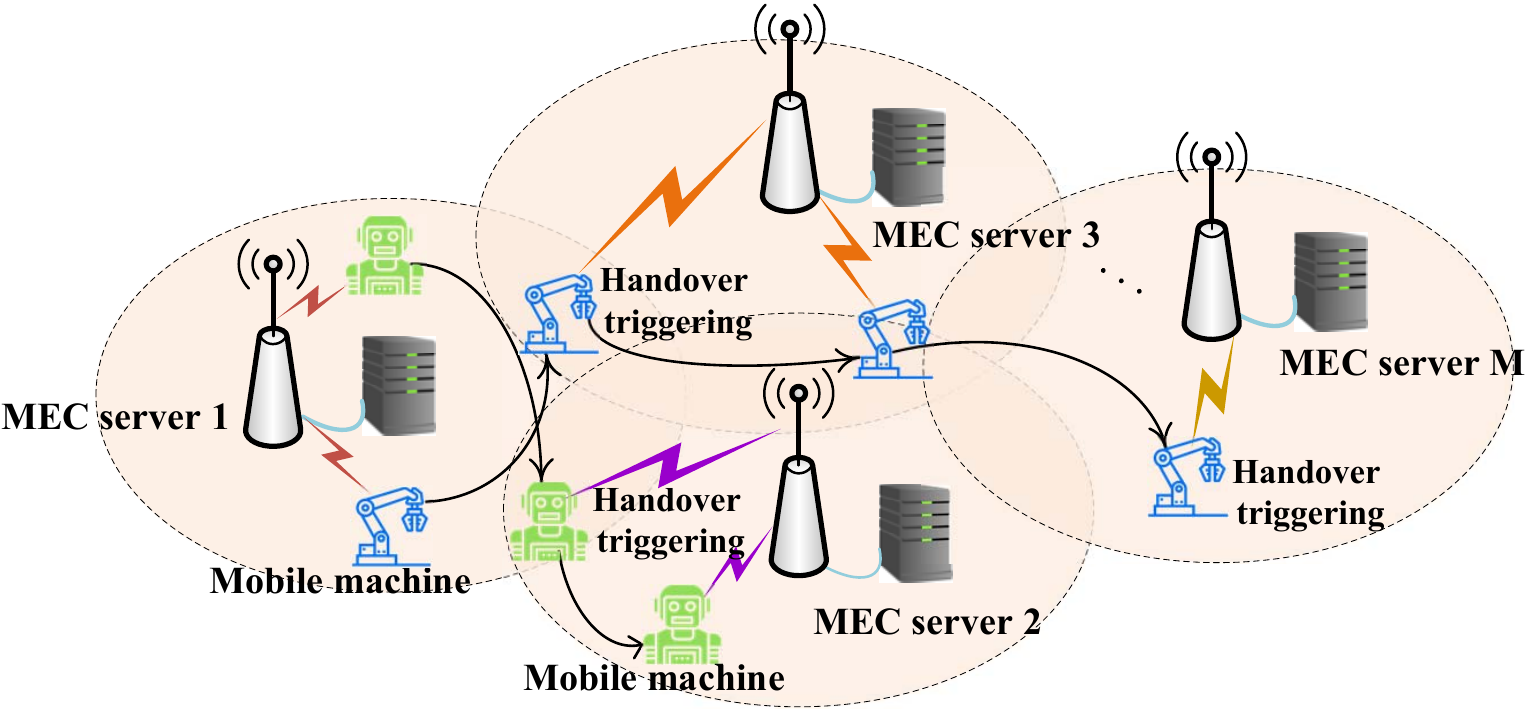}
	\caption{An example of network association system including $M$ MEC servers and a mobile machines in IoT.}
	\label{pic_SysMod}
\end{figure}
\section{System Model}
The network association system in consideration consists of a mobile machine and $M$ small base stations (SBSs), where the corresponding MEC server is deployed at each SBS, and the mobile machine can make proactive network association decision to offload or distribute the computation tasks to the MEC servers for processing. Let $\mathcal{M}=\{1,2,\cdots,M\}$ collect the indexes for $M$ MEC servers. The system model in this paper can be illustrated in Fig. 1.

\subsection{Computational Task Model}
The mobile machine generates $R$ computing tasks to be offloaded during the movement over time. Let ${\cal R} = \{ 0,1, \ldots ,R - 1\}$ denote the set of computational task indexes. For each task $r \in \mathcal{R}$, we define a doublet $\mathcal{X}\buildrel \Delta \over = [{\alpha}(r),{\varepsilon}{\alpha}(r)]$ to parameterize the mobile machine's computation task model, where ${\alpha}(r)$ is the input data size of $r$-th task and $\varepsilon$ represents the computation intensity that indicates the CPU cycles required to process a single computation task. Without loss of generality, we assume that the task size ${\alpha}(r)$ satisfies i.i.d. with unknown probability distribution function. But the task size is constrained by the minimum $A_{\min}$ and the maximum $A_{\max}$.
\subsection{Network Model}
The set of candidate association servers for the $r$-th task of the mobile machine can be defined as 
\begin{equation}
\label{candidateServer}
{{\cal A}}(r) = \left\{ {j|\left\| {{\ell}(r) - {\ell _j}} \right\| \le Ra,\forall j \in {\cal M}} \right\} \,,
\end{equation}
where ${\ell}(r)$ is the position corresponding to the $r$-th task generated by the mobile machine, ${\ell}_{j}$ indicates the deployment location of server $j$ and $Ra$ is the coverage radius of each server. 

Let $c_{j}(r) \in [c_{j}^{\min},c_{j}^{\max}]$ denote the channel capacity between the mobile machine and the MEC server $j$ during the processing of $r$-th task, where $c_{j}^{\min}$, $c_{j}^{\max}$ represent the minimum and the maximum capacity, respectively. Moreover, i.i.d. flat block fading channels are assumed, that is, the channel capacity remains unchanged during the same task processing and varies between different tasks. 
Similarly, we assume that the CPU-cycle frequency allocated by the MEC server $j$ satisfies i.i.d. stochastic process with the minimum $f_{j}^{\min}$ and the maximum $f_{j}^{\max}$, that is, $f_{j}(r) \in [f_{j}^{\min},f_{j}^{\max}]$.  

\subsection{Task Processing Model}
For the computation task $r \in \mathcal{R}$ of the mobile machine, we introduce the following two definitions: 
\begin{itemize}
	\item \textbf{Semi-Markov states:} ${{\bf{s}}}(r) = [{\ell}(r),{m}(r)]$ is the current state of task $r$, including the machine location ${\ell}(r)$ and the previous machine association ${m}(r)$, which can take values from the discrete state space $\mathcal{S}$ with cardinality $\mathrm{S}$. We assume that the task state transfers according to the Markov chain and operates in continuous time.
	\item \textbf{Random events:} ${{\bf{w}}}(r) = [{\alpha}(r),{c_{j}}(r),{f_{j}}(r),\forall j \in {\cal A}(r)]$ is defined as random events for task $r$. Each component of ${{\bf{w}}}(r)$ is i.i.d. over different tasks with an unknown probability distribution.
\end{itemize}

\begin{figure}[!t]
	\centering
	\subfigure[Timeslot-based structure]
	{	\begin{minipage}{0.45\textwidth}
			\centering
			\includegraphics[width=\textwidth]{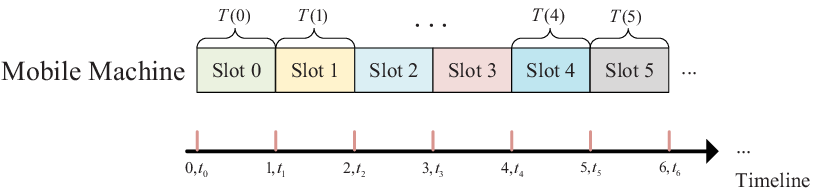}
		\end{minipage}
	}
	\subfigure[Event-triggered structure]
	{	\begin{minipage}{0.45\textwidth}
			\centering
			\includegraphics[width=\textwidth]{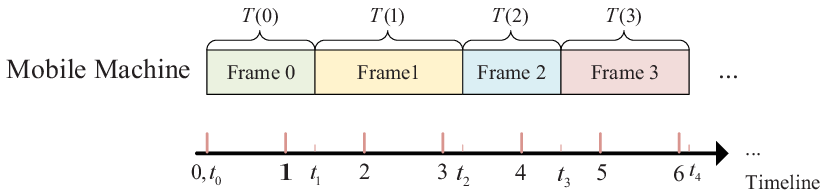}
		\end{minipage}
	}
	\caption{An illustration of the difference between network association decision based on timeslot-based structure and event-triggered structure, where the timeline is divided to back-to-back intervals according to timeslot length and task frame size, respectively.}
	\label{Fig2}
\end{figure}

Note that the above definitions are given from the event-triggered perspective, that is, the event-triggering of network association decision depends on the completion time of computational task, which is different from the traditional slot-based structure of network association model with fixed equal decision intervals, as depicted in Fig. \ref{Fig2}.  
The network handover occurs when the network association $a(r)$ is different from the previous one $m(r)$. Then the handover delay can be given by
\begin{equation}
\label{handoverDelay}
\begin{array}{l}
d_h({{\bf{s}}}(r),{{\bf{w}}}(r),{a}(r))\\
= \left\{ {\begin{array}{*{20}{c}}
	{0,}&{{a}(r) = {m}(r)}\\
	{C,}&{{a}(r) \ne {m}(r),}
	\end{array}} \right.
\end{array}
\end{equation}
where $C$ is the total constant cost of signaling overhead and task migration caused by one handover.

After the association decision ${a}(r)$ is selected, the transmission delay of the computing task offloaded from the mobile machine to the server $a(r)$ can be expressed as 

\begin{equation}
\label{transmissionDelay}
d_{tr}({{\bf{s}}}(r),{{\bf{w}}}(r),{a}(r)) = \frac{{{\alpha}(r)}}{{{c_{{a}(r)}}}(r)} \,,
\end{equation}
then the task computation delay processed by the server $a(r)$ can be given by
\begin{equation}
\label{computationDelay}
d_c({{\bf{s}}}(r),{{\bf{w}}}(r),{a}(r)) = \frac{{{\alpha}(r){\varepsilon}}}{{{f_{{a}(r)}}}(r)} \,.
\end{equation}
According to (\ref{handoverDelay}) - (\ref{computationDelay}), we can define the total processing delay of task $r$ (i.e., task delay) as follows
\begin{equation}
\label{totalDelay}
\begin{split}
{d}(r) & \triangleq d_h({{\bf{s}}}(r),{{\bf{w}}}(r),{a}(r)) + d_{tr}({{\bf{s}}}(r),{{\bf{w}}}(r),{a}(r)) \\
& + d_c({{\bf{s}}}(r),{{\bf{w}}}(r),{a}(r)) \,,
\end{split}
\end{equation}

Moreover, the energy consumption of machine caused by the uplink transmission for computing task can be given by 
\begin{equation}
\label{energyConsumption}
{e}({{\bf{s}}}(r),{{\bf{w}}}(r),{a}(r)) = \frac{{p_{\textrm{tx}}{\alpha}(r)}}{{{c_{{a}(r)}}}(r)} \,,
\end{equation}
where $p_{\rm{tx}}$ is the transmit power of the mobile machine. 

According to the above definition, we can define the task frame size as ${T}(r) \buildrel \Delta \over = {d}(r)$, which means that the adjacent decision interval is equal to the processing delay of previous task under the principle of first-in-first-out (FIFO) execution. 
\newcounter{mytempeqncnt}
\begin{figure*}[hb]
	\hrulefill
	\normalsize
	\setcounter{mytempeqncnt}{\value{equation}}
	\setcounter{equation}{11}
	\begin{equation}
	\label{mobilityDelayCost}
	\begin{array}{l}
	{C}({{\bf{s}}}(r),{{\bf{w}}}(r),{a}(r),{{\bf{s}}}(r + 1))
	= \left\{ {\begin{array}{*{20}{c}}
		{d_{tr}(r) + d_c(r),}&{{{a}}(r) = {m}(r),{{a}}(r) \in {{\cal A}}(r + 1)}\\
		{d_{tr}(r) + d_c(r) + d_h(r),}&{{{a}}(r) = {m}(r),{{a}}(r) \notin {{\cal A}}(r + 1)}\\
		{d_{tr}(r) + d_c(r) + d_h(r),}&{{{a}}(r) \ne {m}(r),{{a}}(r) \in {{\cal A}}(r + 1)}\\
		{d_{tr}(r) + d_c(r) + 2d_h(r),}&{{{a}}(r) \ne {m}(r),{{a}}(r) \notin {{\cal A}}(r + 1)} \,.
		\end{array}} \right.
	\end{array}
	\end{equation}
\end{figure*}
\section{Two-stage Online Proactive network Association decision}
We first define the frame average of task processing delay (i.e., the average task delay) as follows 
\begin{equation}
\label{frameAvgDelay}
{\overline d}(R) = \frac{1}{R}\sum\nolimits_{r = 0}^{R - 1} \mathop{\mathbb{E}}{[{d}(r)]} \,.
\end{equation}
 Similarly, the frame average of energy consumption and task frame size can be defined as ${\overline e}(R)$, ${\overline T}(R)$, respectively. Then the time average of energy consumption can be given by 
\begin{equation}
\label{timeAvgEnergyConsump}
\frac{{{{\overline e }}}}{{{{\overline T }}}} = \mathop {\lim }\limits_{R \to \infty } \frac{{\frac{1}{R}\sum\nolimits_{r = 0}^{R - 1} \mathbb{E}{[{e}(r)]} }}{{\frac{1}{R}\sum\nolimits_{r = 0}^{R - 1} \mathbb{E}{[{T}(r)]} }} \,,
\end{equation}
where we rewrite $e(r)={e}({{\bf{s}}}(r),{{\bf{w}}}(r),{a}(r))$ for simplicity of notation, and we assume that the above stochastic processes ${d}(r)$, ${e}(r)$ and ${T}(r)$ converge to ${\overline d}$, ${\overline e}$ and ${\overline T}$ as the number of computation tasks $R \rightarrow \infty$. Then we can formulate $\textbf{P}$ to minimize the average task delay as
\begin{equation}
\label{optimizationP}
\begin{aligned}
{\textbf{P}}:\quad &\min \;\;{{\overline d }} \\
&{\textbf{C1}}: {\frac{{{{\overline e }}}}{{{{\overline T }}}} \le \beta } \,; \quad{\textbf{C2}}:{{a}}(r) \in {{\cal A}}(r),\;\forall r \in {\cal R} \,,
\end{aligned}
\end{equation}
where $\beta$ is a given constant to specify the upper bound of time-averaged energy consumption constraint. Note that the above optimization problem is difficult to solve due to the existences of semi-Markov task state and random events. More specifically, the traditional MDP approach is inapplicable due to the unknown state transition probability and the curse of dimensionality caused by random events with unknown probability distribution. Furthermore, this problem is not a standard stochastic optimization due to the coupling introduced by the semi-Markov task states.
\subsection{Semi-Markovian Model and Mobility-aware Anticipatory Association}
We first introduce a semi-Markovian model to precisely describe the behavior of task state. The task state transition probability can be derived as follows
\setcounter{mytempeqncnt}{\value{equation}}
\setcounter{equation}{9}
\begin{equation}
\label{trProbability}
\begin{split}
& \Pr \{ {{\bf{s}}}(r + 1)|{{\bf{s}}}(r),{a}(r)\} \\
=& \Pr ({m}(r + 1)|{m}(r),{a}(r)) \cdot \Pr({\ell}(r + 1)|{\ell}(r)) \,,
\end{split}
\end{equation}
where the conditional transition probability of network association satisfies
\begin{equation}
\label{associationPrtr}
\Pr ({m}(r + 1)|{m}(r),{a}(r)) = \left\{ {\begin{array}{*{20}{c}}
	{1,}&{{\textrm{if}}\;{m}(r + 1) = {{a}}(r)}\\
	{0,}&{{\textrm{otherwise}}.}
	\end{array}} \right.
\end{equation}

Considering the possibility that the mobile machine's future mobility trajectory triggers a potential network handover, we define the task delay cost as given by (\ref{mobilityDelayCost}). Then the expected task delay cost can be given by 
\setcounter{mytempeqncnt}{\value{equation}}
\setcounter{equation}{12}
\begin{equation}
\label{expectedDelayCost}
\begin{split}
& d^{'}({{\bf{s}}}(r),{{\bf{w}}}(r),{a}(r)) \\
&= \sum\nolimits_{\mathbf{s}(r + 1) \in {\mathcal S}} C\left( {{\bf{s}}}(r),{{\bf{w}}}(r),{a}(r),{{\bf{s}}}(r + 1) \right) \\
& \qquad \qquad \quad \quad \cdot Pr({{\bf{s}}}(r + 1)|{{\bf{s}}}(r),{a}(r)) \,.
\end{split}
\end{equation}
As a result, we construct the mobility-aware anticipatory network association problem \textbf{P1} as follows
\begin{equation}
\label{optimizationP1}
\begin{aligned}
{\textbf{P1}}:\quad &\min \quad \overline{d^{'}} \\
&{\textbf{C1}}:\frac{\overline{e}}{{\overline T }} \le \beta\,; \quad {\textbf{C2}}:{{a}}(r) \in {{\cal A}}(r),\;\forall r \in {\cal R} \,.
\end{aligned}
\end{equation}

\subsection{The Calculation of Optimal Proactive Network Association Parameters}
Considering the coupling of task state, we first treat it as the decision variable rather than an observation of the environment to find the optimal parameters of proactive network association decision \cite{neely2011online}. We introduce the following attribute as the observation during task frame, which suggests
\begin{equation}
\label{trAttribute}
q_{mn}(r) = 1({\mathbf{s}}(r) = m)\Pr ({\mathbf{s}}(r + 1) = n|{\mathbf{s}}(r) = m,{a}(r)) \,,
\end{equation}
where $1({\mathbf{s}}(r) = m)$ is an indicator function. Then we can reformulate optimization problem as follows
\begin{equation}
\label{optimizationP2}
\begin{aligned}
{\textbf{P2}}:\quad &\min \quad \overline {d^{'}} \\
\textrm{s.t.} \; &{\textbf{C1}}:\;\frac{{\overline {{e}} }}{{\overline {{T}} }} \le {\beta}\,;  {\textbf{C2}}:\sum\nolimits_{m = 1}^S {\overline {q_{sm}}  = } \sum\nolimits_{n = 1}^S {\overline {q_{ns}} ,\forall s \in {\cal S} } \,;\\
&{\textbf{C3}}:\;{{\bf{s}}}(r) \in {\cal S},\;\forall r \in {\cal R}\,; \;{\textbf{C4}}:\;{{a}}(r) \in {{\cal A}}(r),\forall r \in {\cal R}\,;\\
&{\textbf{C5}}:\;{{\bf{s}}}(r)\;{\textrm{is}}\;{\textrm{selected}}\;{\textrm{independently}}\;{\textrm{of}}\;{{\bf{w}}}(r),
\end{aligned}
\end{equation}
where constraints \textbf{C1}, \textbf{C3} and \textbf{C4} are self-explanatory. \textbf{C2} ensures the global balance of state transition, which specifies that the frame average of task state in $s$ is equal to the frame average of transitioning to state $s$, Furthermore, \textbf{C5} ensures that the decision variable selection of the task state should follow the principle of independence from random events. 

According to the Lyapunov optimization theory \cite{neely2010stochastic}, we define a virtual queue of energy consumption deviation, which can be evolved as 
\begin{equation}
\label{energyVirtualQ}
{E}(r + 1) = \max\{ {E}(r) + {e}(r) - {\beta}{T}(r),0\} \,.
\end{equation}
Similarly, we can define virtual queue $G_s(r)$ to ensure the frame average equality constraint \textbf{C2}, which is updated by
\begin{equation}
\label{trVirtualQ}
G_s(r + 1) = G_s(r) + 1({\mathbf{s}}(r) = s) - \sum\nolimits_{n = 1}^S {q_{ns}(r)} \,.
\end{equation}
Then the Lyapunov function can be defined as 
\begin{equation}
\label{lyaF}
L(r) = \frac{1}{2}{E}{(r)^2} + \frac{1}{2}\sum\nolimits_{s = 1}^S {G_s{{(r)}^2}} \,,
\end{equation}
and the drift-plus-penalty function can be given by 
\begin{equation}
\label{driftplusPenalty}
{\Delta _V}(r) = \Delta (r) + V\mathbb{E}[d^{'}(r)|\Theta(r)] \,,
\end{equation}
where ${\Theta}(r) = \{ {E}(r),G_1(r),G_2(r),...,G_S(r)\}$ is the collection of all queue backlogs at task frame $r$, and $\Delta (r) \triangleq \mathbb{E}[L(r + 1) - L(r)|\Theta(r)]$ is the conditional Lyapunov drift. Moreover, $V$ is a predefined Lyapunov parameter. Then the upper bound of (\ref{driftplusPenalty}) can be derived in the following lemma.
\begin{lemma}
	At task frame $r$, the drift-plus-penalty function (\ref{driftplusPenalty}) satisfies the following inequality:
	\begin{equation}
	\label{upperboundDPP}
	\begin{aligned}
	&{\Delta _V}(r) \le B + V\mathbb{E}[d^{'}(r)|\Theta(r)] + {E}(r)\mathbb{E}[{e}(r) - {\beta}{T}(r)|\Theta(r)]\\
	&+ \sum\nolimits_{s = 1}^S {G_s(r)}\mathbb{E}[1({{\bf{s}}}(r) = s) - \Pr (s|{{\bf{s}}}(r),{a}(r))|\Theta(r)] \,,
	\end{aligned}
	\end{equation}
	where $B$ is a finite constant and can be defined in the proof.
\end{lemma}

The proof is given in Appendix. Inspired by the maximum weight learning algorithm developed in \cite{neely2012max}, we use $W$ latest past samples of random events to evaluate the expectation value in task state $s$ instead of the exact probability distribution knowledge, which can be given by 
\begin{equation}
\label{sampleApproxima}
\begin{aligned}
{\hat e_s}(r) = \frac{1}{W}\sum\limits_{w = 1}^W {\mathop {\min }_{{a}^{(w)}(r) \in {{\cal A}}(r)} } & [p({{\bf{s}}}(r) = s,\\
& {\bf{w}}^{(w)}(r),{a}^{(w)}(r))|\Theta (r)] \,,
\end{aligned}
\end{equation}
where $p({{\bf{s}}}(r) = s,{\bf{w}}^{(w)}(r),{a}^{(w)}(r))$ is the simplified expression of the right-hand side of (\ref{upperboundDPP}) after the expectation and constant $B$ are removed. Then the task state $\mathbf{s}^{*}(r)$ can be selected as the minimizer of ${\hat e_s}(r)$, $\forall s \in \mathcal{S}$.

After the task state $\mathbf{s}^{*}(r)$ is selected, the mobile machine observes the exact value of random events at current task frame $r$, and then the network association decision can be made to minimize the deterministic minimization problem of (\ref{upperboundDPP}), which can be given by 
\begin{equation}
\label{determiAssociaF}
\begin{aligned}
&\min\limits_{{{a}}(r) \in {{\cal A}}(r)} [Vd^{'}(r) + {E}(r)({e}(r) - {\beta}(r){T}(r)) \\
& + \sum\nolimits_{s = 1}^S {G_s(r)(1({{\bf{s}}}(r) = s) - Pr(s|{{\bf{s}}}(r),{a}(r)))} ] \,.
\end{aligned}
\end{equation} 
Moreover, the optimal parameters of energy consumption, task frame size and state transition probability can be given by
\begin{equation}
\label{optimalEnergyCost}
\mathrm{e}_{s}^{*} = \frac{{\overline {1({{\bf{s}}^{*}}(r) = s){e}(r)} }}{{\overline {1({{\bf{s}}^{*}}(r) = s)} }} \,,
\end{equation}
\begin{equation}
\label{optimalFrameSize}
\mathrm{T}_{s}^{*} = \frac{{\overline {1({{\bf{s}}^{*}}(r) = s){T}(r)} }}{{\overline {1({{\bf{s}}^{*}}(r) = s)} }} \,,
\end{equation}
\begin{equation}
\label{optimalTrP}
\mathrm{Pr}_{sm}^{*} = \frac{{\overline {1({{\bf{s}}^{*}}(r) = s)q_{sm}(r)} }}{{\overline {1({{\bf{s}}^{*}}(r) = s)} }} \,,
\end{equation}
\subsection{The Online Proactive Network Association Decision} 
Under the guidance of optimal parameters obtained by (\ref{optimalEnergyCost})-(\ref{optimalTrP}), the following optimization problem can be constructed equivalently as follows
\begin{equation}
\label{optimizationP3}
\begin{aligned}
{\bf{P3}}: \quad &\min \;\;\;\;\overline {d^{'}} \\
\textrm{s.t.}\quad &{\bf{C1}}:\;\overline {{e}}  \le \sum\nolimits_{s = 1}^S {\overline {1({{\bf{s}}}(r) = s)} } \mathrm{e}_{s}^{*},\;\forall r \in {\cal R}\\
&{\bf{C2}}:\;\overline {{T}}  = \sum\nolimits_{s = 1}^S {\overline {1({{\bf{s}}}(r) = s)} } \mathrm{T}_{s}^{*},\;\forall r \in {\cal R}\\
&{\bf{C3}}:\;\overline {q_{sm}}  = \overline {1({{\bf{s}}}(r) = s)} \mathrm{Pr}_{sm}^{*},\;\forall r \in {\cal R}\\
&{\bf{C4}}:\;{{a}}(r) \in {{\cal A}}(r),\;\forall r \in {\cal R} \,.
\end{aligned}
\end{equation}
where the task state is observed from network environment instead of being determined. Moreover, constraints \textbf{C1} and \textbf{C2} can jointly ensure that the time-averaged energy consumption of \textbf{P1} is satisfied. Constraint \textbf{C3} indicates that the frame fraction of state transition is equal to the optimal state transition probability. Similarly, the virtual queues can be introduced as follows
\begin{equation}
\label{energyVirtualQ2}
{\hat{E}}(r + 1) = \max [{\hat E}(r) + {e}(r) - \sum\nolimits_{s = 1}^S {1({{\bf{s}}}(r) = s)\mathrm{e}_{s}^{*}},0] \,,
\end{equation}
\begin{equation}
\label{frameSizeVirtualQ}
{\hat F}(r + 1) = {\hat F}(r) + {T}(r) - \sum\nolimits_{s = 1}^S 1({{\bf{s}}}(r) = s)\mathrm{T}_{s}^{*} \,,
\end{equation}
\begin{equation}
\label{trVirtualQ2}
\hat G_{sm}(r + 1) = \hat G_{sm}(r) + q_{sm}(r) - 1({\bf{s}}(r) = s)\mathrm{Pr}_{sm}^{*} \,.
\end{equation}
Define the drift-plus-penalty function as 
\begin{equation}
\label{driftPlusPenalty2}
{\hat \Delta _{\hat V}}(r) = \hat \Delta (r) + \hat V\mathbb{E}[d^{'}(r)|\hat \Theta(r)] \,,
\end{equation}
where $\hat \Theta(r) = \{ {\hat E}(r),{\hat F}(r),\hat G_{sm}(r),\forall s,m \in {\cal S}\}$ collects all virtual queues, $\hat \Delta (r)$ is conditional Lyapunov drift and $\hat V$ is the Lyapunov control parameter. Then the upper bound of (\ref{driftPlusPenalty2}) can be derived in Lemma 2.
\begin{lemma}
	 The Lyapunov drift-plus-penalty function (\ref{driftPlusPenalty2}) satisfies as follows:
	\begin{equation}
	\label{DPPUpperbound2}
	\begin{aligned}
	&{{\hat \Delta }_{\hat V}}(r) \le \hat B + \hat V[d^{'}(r)|\hat \Theta(r)] + {{\hat E}}(r)\mathbb{E}[{e}(r)\\
	& - \sum\nolimits_{s = 1}^S {1({{\bf{s}}}(r) = s)e_{s}^{*}|\hat \Theta(r)} ] \\
	&+ {\hat F}(r)\mathbb{E}[{T}(r) - \sum\limits_{s = 1}^S {1({{\bf{s}}}(r) = s)T_{s}^{*}|\hat \Theta(r)} ] +\\
	&\sum\nolimits_{s = 1}^S {\sum\nolimits_{m = 1}^S {\hat G_{sm}(r)\mathbb{E}[{q}_{sm}(r) - 1({s}(r) = s)\mathrm{Pr}_{sm}^{*}|\hat \Theta(r)]} } \,,
	\end{aligned}
	\end{equation}
	where $\hat{B}$ is a finite constant. 
\end{lemma}

The proof process is omitted due to the similarity to Lemma 1. Then the proactive network association can be performed to solve the deterministic optimization problem in right-hand side of (\ref{DPPUpperbound2}) at each task frame. Moreover, the detailed procedure of TOPNA to execute online proactive network association decision can be described in Algorithm 1.
\begin{algorithm}[htbp]
	\caption{An online proactive algorithm to execute network association for mobile machines}
	\label{Alg2}
	\begin{algorithmic}[1]
		\STATE Initialize the optimal association parameter $\mathrm{e}_{s}^{*}$, $\mathrm{T}_{s}^{*}$ and $\mathrm{Pr}_{sm}^{*}$ according to (\ref{optimalEnergyCost}), (\ref{optimalFrameSize}) and (\ref{optimalTrP}).
		\REQUIRE
		\STATE Obtain the task state ${{\bf{s}}}(r)$ and observe random event ${{\bf{w}}}(r)$, virtual queues ${\hat E}(r)$, ${\hat F}(r)$ and $\hat G_{sm}(r)$.
		\STATE Determine the network association decision ${a}(r)$ by minimizing the deterministic optimization of (\ref{DPPUpperbound2}).
		\STATE Update ${\hat E}(r + 1)$, ${\hat F}(r + 1)$ and $\hat G_{sm}(r + 1)$ based on (\ref{energyVirtualQ2}), (\ref{frameSizeVirtualQ}) and (\ref{trVirtualQ2}), respectively.
	\end{algorithmic}
\end{algorithm}	
\section{Simulation Results}
In this section, we compare the performance of TOPNA with other three network association policies, that is, ``Best channel", ``Max-sojourn" and ``Myopic optimal" policies indicate that the mobile machine selects MEC server within candidate association set according to the best channel quality, the maximum sojourn time and the minimum delay relative to transmission and computation, respectively.

In our simulation, the MEC servers are deployed on a regular grid network within a square area of $1000 \times {\rm{1000}}\;{{\rm{m}}^2}$, and the coverage radius of server $Ra$ is $200 \rm{m}$. The mobile machine's trajectory is taken from real-world GeoLife dataset \cite{zheng2009mining}. The transmit power of mobile machine is $p_\textrm{tx}= 23\;\textrm{dbm}$. The computation intensity of task is ${\varepsilon} = 238\;{\rm{CPU\;cycles/bit}}$. The number of past samples for random events is $W=50$. Specifically, the task data size satisfies ${\alpha}(r)\sim U(0.5,1)\;\textrm{Mbit}$, where $U(a,b)$ denotes the uniform distribution with parameters $a$ and $b$. Moreover, we assume that the average channel capacity $c_j$, $\forall j \in \mathcal{M}$ takes values from $[20,100]\;\textrm{Mbps}$ with evenly spaced intervals, then the channel capacity of each task frame satisfies ${c_{j}}(r)\sim U(0.5{c_j},1.5{c_j})$. Similarly, we assume that the average server's CPU-cycle frequency ${f_j}$, $j \in \mathcal{M}$ takes values from $[10,50]\;\textrm{GHz}$ with evenly intervals, then the available CPU-cycle frequency for mobile machine ${f_{j}}(r)\sim U(0.5{f_j},1.5{f_j})$. Furthermore, the Lyapunov control parameters of TOPNA are set to the same value, i.e., $V=\hat{V}$. Each experiment is run in MATLAB for $3000$ task frames. 

\begin{figure}[t]
	\centering
	\includegraphics[width=0.38\textwidth]{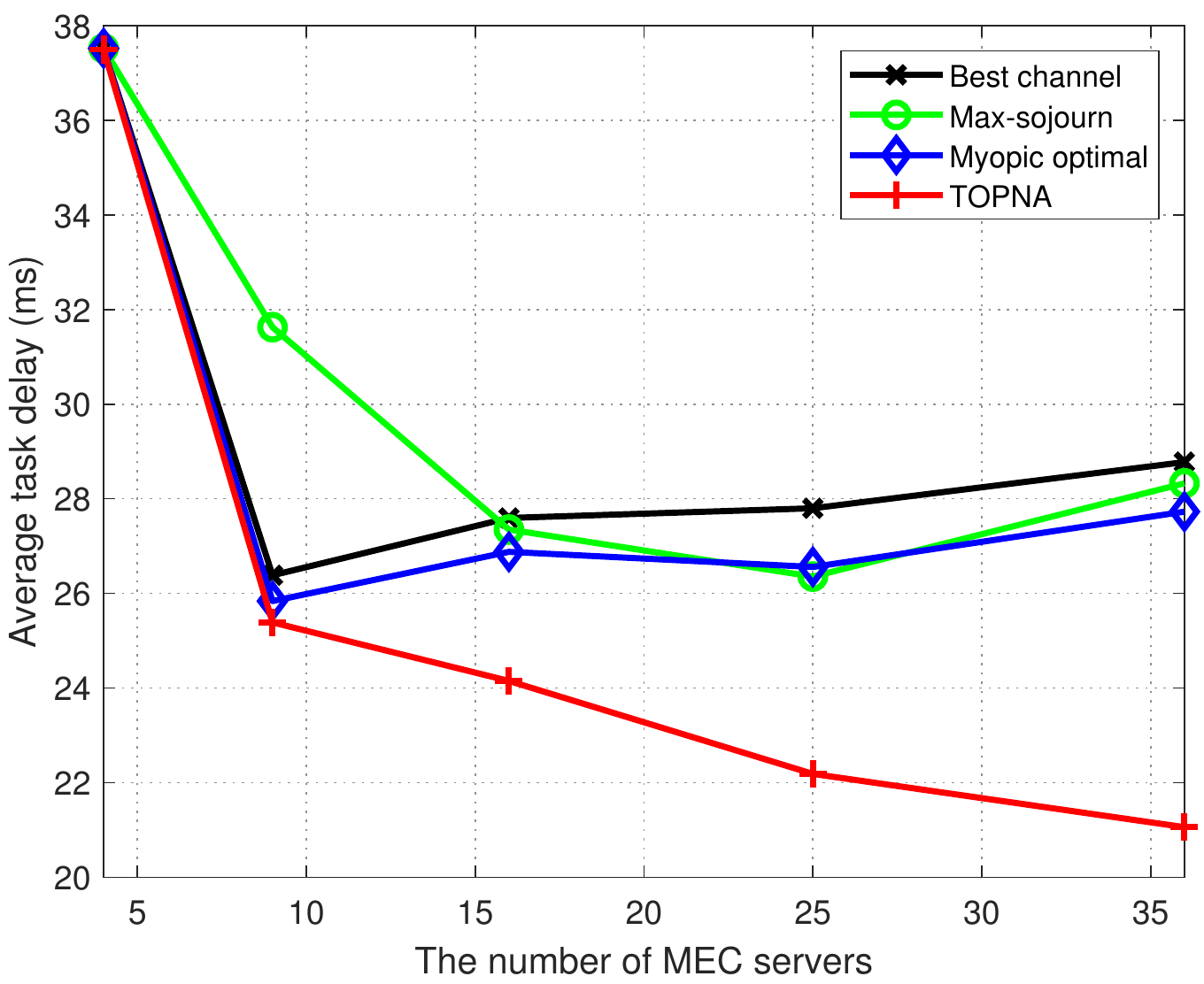}
	\caption{The changes of average task delay and the corresponding number of MEC servers as $M$ increases from $4$ to $36$, where $V=500$, $C=15\;\textrm{ms}$ and $\beta = 125\;\textrm{mJ/s}$.}
	\label{Fig3}
\end{figure}

In Fig. \ref{Fig3}, we plot the average task delay against the number of MEC servers for proposed TOPNA and other comparison schemes. We obtain that the average task delay of each scheme coincides exactly when $M=4$, this is due to no overlapping coverage of MEC servers, that is, the candidate server is unique for each task. As $M$ increases, the average task delay of proposed TOPNA algorithm gradually decreases due to the increased chance to associate the optimal server through more overlapping coverage areas, and is lower than other schemes by comprehensively considering the impacts of transmission, computation, handover and future mobility on task delay. As a result, it shows that our propose TOPNA algorithm performs more delay-efficient than other three comparison schemes, especially in dense deployment of MEC server.
\begin{figure}[!t]
	\centering
\subfigure[Task delay]
{	\begin{minipage}{0.38\textwidth}
		\centering
		\includegraphics[width=\textwidth]{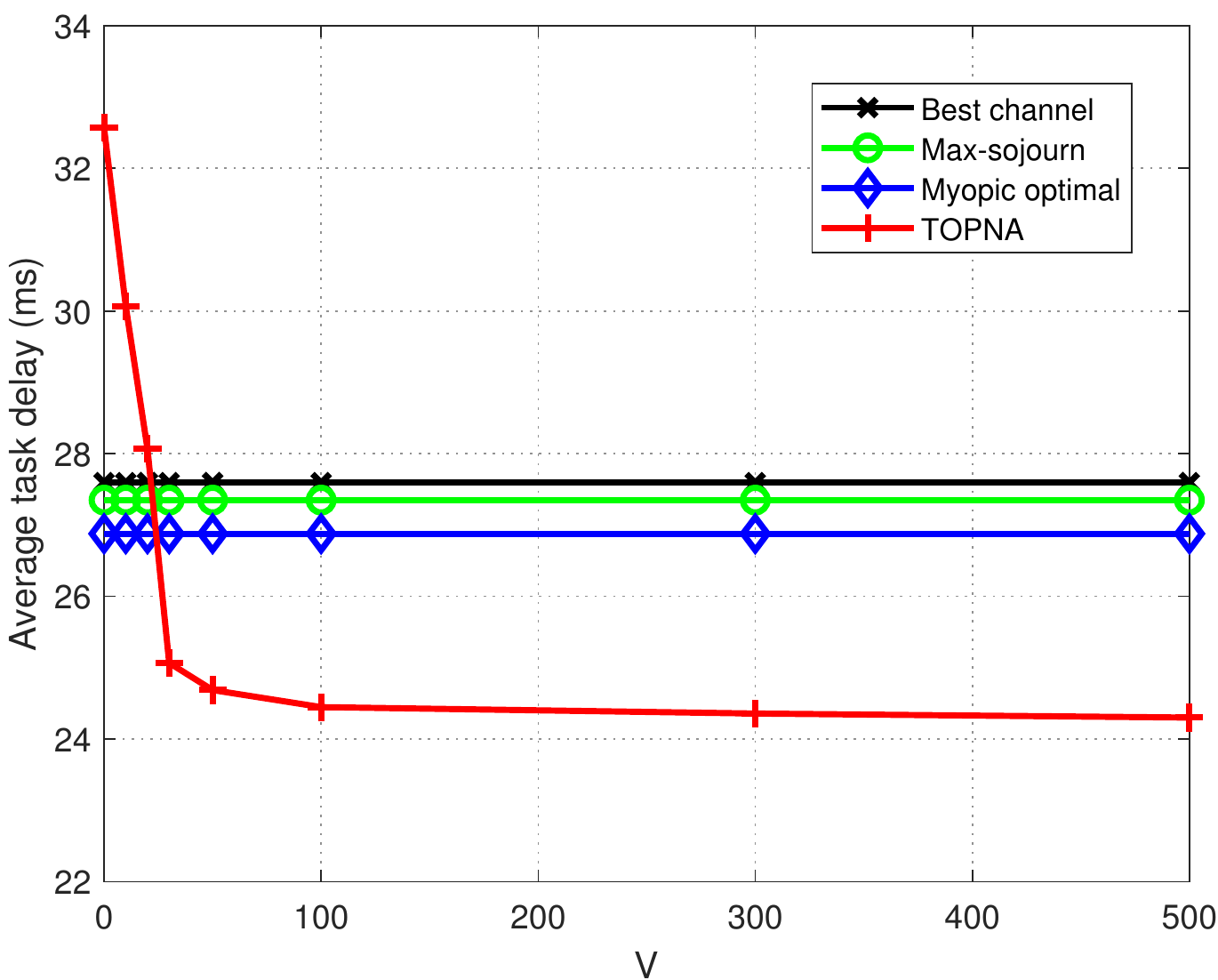}
	\end{minipage}
}
\subfigure[Task energy consumption]
{	\begin{minipage}{0.38\textwidth}
		\centering
		\includegraphics[width=\textwidth]{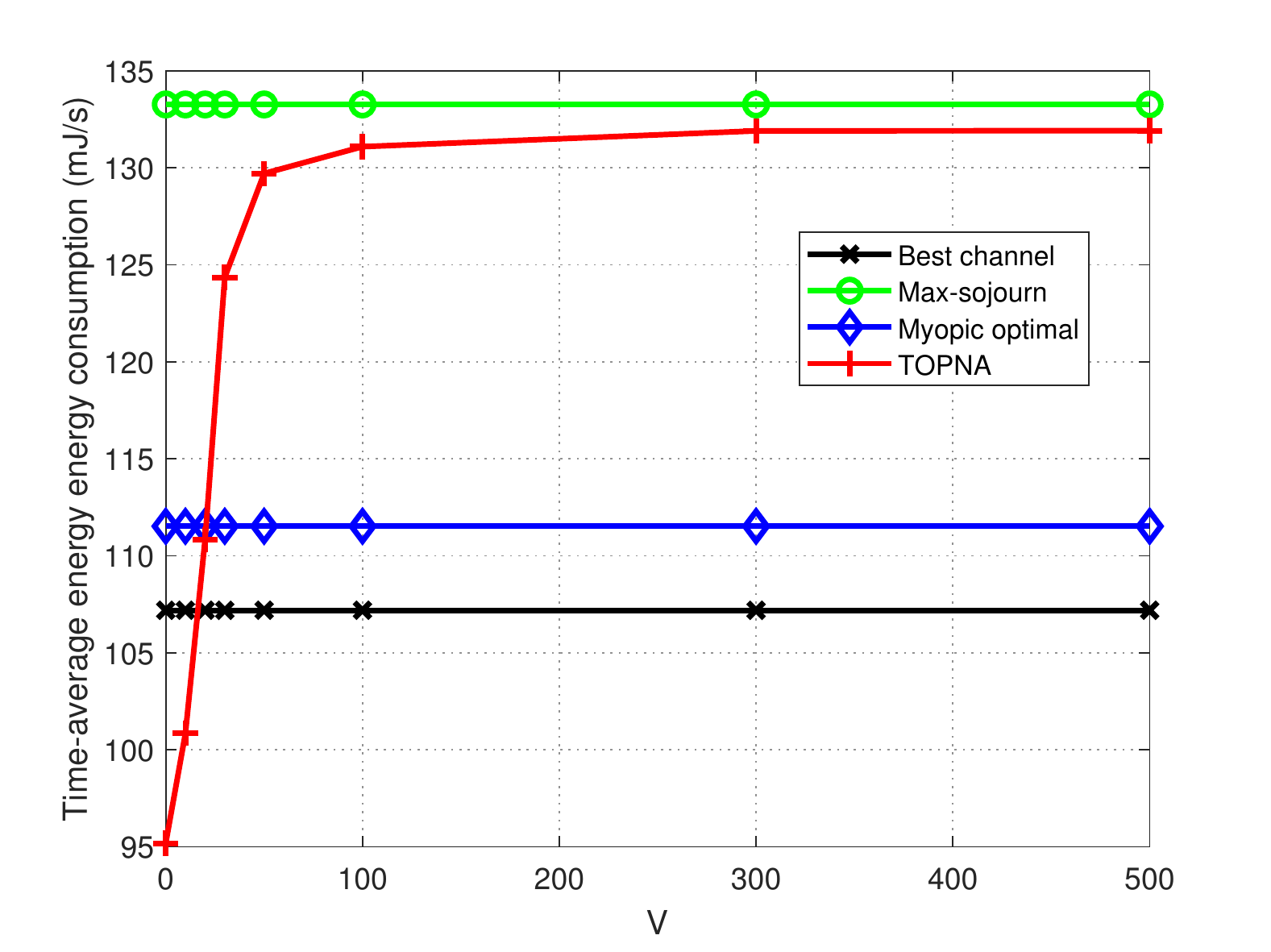}
	\end{minipage}
}
\caption{The average task delay and time-average task energy consumption of the proposed TOPNA algorithm and other three schemes, where V ranges from $0$ to $500$, $M=16$, $C=15\;\textrm{ms}$, and $\beta=125\; \textrm{mJ/s}$.}
\label{Fig4}
\end{figure}

Fig. \ref{Fig4} shows the average task delay and time-average energy consumption of proposed TOPNA method and other three schemes with increasing $V$. We find that, when control parameter $V\leq50$, the average task delay and time-average energy consumption accordingly decrease and increase rapidly, and then tend to be stable due to the limited channel capacity and server CPU-cycle frequency when $V\geq100$. Meanwhile, the average task delay of the proposed TOPNA satisfies the asymptotic optimality with bounded deviation and has a trade-off relationship with the time-average energy consumption. Besides, we also plot the performance of other three schemes, each of which is independent of $V$ and therefore remains unchanged. Moreover, the simulation results provide the principle that the control parameter $V$ is selected in practice, that is, under the energy consumption budget, selecting the appropriate $V$ can minimize the average task delay performance.

\section{Conclusion}
In this paper, we propose an online proactive network association method based on event-triggered distributed computing environment, to minimize the average task delay subjected to time-averaged energy consumption for mobile machines in IoT under the assistance of mobile edge computing. We first formulate an event-triggered computing and network association model by taking into account semi-Markov task states and i.i.d. random events. Then a mobility-aware anticipatory association is developed to consider the handover caused by mobile machine's mobility. Finally, the two-stage online proactive network association (TOPNA) decision algorithm based on the MDP and Lyapunov optimization is proposed. Simulation results demonstrate the effectiveness of our proposed algorithm. Future research directions include establishing a more general method of proactive network association for multiple machines in distributing computing and networking scenario. 

\appendices
\section{}
For any positive real number $a$, $b$, $c$, the following inequality holds:
\begin{equation}
\label{inequaFormu}
{(\max[a - b,0] + c)^2} \le {a^2} + {b^2} + {c^2} + 2a(c - b) \,.
\end{equation}
According to the above property (\ref{inequaFormu}), and taking squares of both sides of (\ref{energyVirtualQ}) and (\ref{trVirtualQ}), we have
\begin{equation}
\label{energyVirtuQIneq}
{E}{(r + 1)^2} - {E}{(r)^2} \le {{\rm{e}}}{(r)^2} + {\beta}^2{T}{(r)^2} + 2{E}(r)[{e}(r) - {\beta}{T}(r)] \,.
\end{equation}
Similarly, we have
\begin{equation}
\label{trVIrtualQIneq}
\begin{aligned}
G_s{(r + 1)^2}& - G_s{(r)^2} \le 1{({{\bf{s}}}(r) = s)^2} + {(\sum\nolimits_{n = 1}^S {q_{ns}(r)} )^2}\\
&+ 2G_s(r)[1({{\bf{s}}}(r) = s) - \sum\nolimits_{n = 1}^S {q_{ns}(r)} ] \,.
\end{aligned}
\end{equation}
Summing up (\ref{energyVirtuQIneq}) and (\ref{trVIrtualQIneq}), then taking expectation on both sides of and plugging into (\ref{driftplusPenalty}), we can obtain the desired result in (\ref{upperboundDPP}). And the constant $B$ satisfies the following:
\begin{equation}
\label{constantBDefi}
B = \frac{1}{2}(e_{\max }^2 + {\beta}^2T_{\max }^2) + S \,,
\end{equation}
where $e_{\max}$ and $T_{\max}$ represent the upper bound of energy consumption and task frame size, respectively. 
\bibliographystyle{IEEEtran}
\bibliography{ProactiveNetworkAssociation}

\begin{thebibliography}{10}
\providecommand{\url}[1]{#1}
\csname url@samestyle\endcsname
\providecommand{\newblock}{\relax}
\providecommand{\bibinfo}[2]{#2}
\providecommand{\BIBentrySTDinterwordspacing}{\spaceskip=0pt\relax}
\providecommand{\BIBentryALTinterwordstretchfactor}{4}
\providecommand{\BIBentryALTinterwordspacing}{\spaceskip=\fontdimen2\font plus
\BIBentryALTinterwordstretchfactor\fontdimen3\font minus
  \fontdimen4\font\relax}
\providecommand{\BIBforeignlanguage}[2]{{%
\expandafter\ifx\csname l@#1\endcsname\relax
\typeout{** WARNING: IEEEtran.bst: No hyphenation pattern has been}%
\typeout{** loaded for the language `#1'. Using the pattern for}%
\typeout{** the default language instead.}%
\else
\language=\csname l@#1\endcsname
\fi
#2}}
\providecommand{\BIBdecl}{\relax}
\BIBdecl

\bibitem{ckc2018ultra}
K.-C. {Chen}, T.~{Zhang}, R.~D. {Gitlin}, and G.~{Fettweis}, ``Ultra-low
  latency mobile networking,'' \emph{IEEE Network}, vol.~33, no.~2, pp.
  181--187, Mar. 2019.

\bibitem{hu2015mobile}
Y.~C. Hu, M.~Patel, D.~Sabella, N.~Sprecher, and V.~Young, ``Mobile edge
  computing—a key technology towards 5g,'' \emph{ETSI white paper}, vol.~11,
  no.~11, pp. 1--16, Sep. 2015.

\bibitem{hung2018delay}
S.~{Hung}, H.~{Hsu}, S.~{Cheng}, Q.~{Cui}, and K.-C. {Chen}, ``Delay guaranteed
  network association for mobile machines in heterogeneous cloud radio access
  network,'' \emph{IEEE Transactions on Mobile Computing}, vol.~17, no.~12, pp.
  2744--2760, Dec. 2018.

\bibitem{wang2016joint}
N.~Wang, E.~Hossain, and V.~K. Bhargava, ``Joint downlink cell association and
  bandwidth allocation for wireless backhauling in two-tier hetnets with
  large-scale antenna arrays,'' \emph{IEEE Transactions on Wireless
  Communications}, vol.~15, no.~5, pp. 3251--3268, May. 2016.

\bibitem{ye2013user}
Q.~Ye, B.~Rong, Y.~Chen, M.~Al-Shalash, C.~Caramanis, and J.~G. Andrews, ``User
  association for load balancing in heterogeneous cellular networks,''
  \emph{IEEE Transactions on Wireless Communications}, vol.~12, no.~6, pp.
  2706--2716, Jun. 2013.

\bibitem{trestian2012game}
R.~Trestian, O.~Ormond, and G.-M. Muntean, ``Game theory-based network
  selection: Solutions and challenges,'' \emph{IEEE Communications surveys \&
  tutorials}, vol.~14, no.~4, pp. 1212--1231, Fourthquarter. 2012.

\bibitem{wang2016network}
J.~Wang, C.~Jiang, Z.~Han, Y.~Ren, and L.~Hanzo, ``Network association
  strategies for an energy harvesting aided super-wifi network relying on
  measured solar activity.'' \emph{IEEE Journal on Selected Areas in
  Communications}, vol.~34, no.~12, pp. 3785--3797, Dec. 2016.

\bibitem{ko2018spatial}
H.~Ko, J.~Lee, and S.~Pack, ``Spatial and temporal computation offloading
  decision algorithm in edge cloud-enabled heterogeneous networks,'' \emph{IEEE
  Access}, vol.~6, pp. 18\,920--18\,932, Mar. 2018.

\bibitem{li2016joint}
Y.~Li, M.~Sheng, Y.~Sun, and Y.~Shi, ``Joint optimization of bs operation, user
  association, subcarrier assignment, and power allocation for energy-efficient
  hetnets,'' \emph{IEEE Journal on Selected Areas in Communications}, vol.~34,
  no.~12, pp. 3339--3353, Dec. 2016.

\bibitem{lyu2017optimal}
X.~Lyu, W.~Ni, H.~Tian, R.~P. Liu, X.~Wang, G.~B. Giannakis, and A.~Paulraj,
  ``Optimal schedule of mobile edge computing for internet of things using
  partial information,'' \emph{IEEE Journal on Selected Areas in
  Communications}, vol.~35, no.~11, pp. 2606--2615, Nov. 2017.

\bibitem{heemels2012introduction}
W.~P. M.~H. {Heemels}, K.~H. {Johansson}, and P.~{Tabuada}, ``An introduction
  to event-triggered and self-triggered control,'' in \emph{2012 IEEE 51st IEEE
  Conference on Decision and Control (CDC)}, Dec. 2012, pp. 3270--3285.

\bibitem{li2017distributed}
S.~Li, Q.~Yu, M.~A. Maddah-Ali, and A.~S. Avestimehr, ``A scalable framework
  for wireless distributed computing,'' \emph{IEEE/ACM Transactions on
  Networking}, vol.~25, no.~5, pp. 2643--2654, Oct 2017.

\bibitem{neely2011online}
M.~J. Neely, ``Online fractional programming for markov decision systems,'' in
  \emph{2011 49th Annual Allerton Conference on Communication, Control, and
  Computing (Allerton)}, Sep. 2011, pp. 353--360.

\bibitem{neely2010stochastic}
M.~J. Neely, ``Stochastic network optimization with application to
  communication and queueing systems,'' \emph{Synthesis Lectures on
  Communication Networks}, vol.~3, no.~1, pp. 1--211, 2010.

\bibitem{neely2012max}
M.~J. Neely, S.~T. Rager, and T.~F. La~Porta, ``Max weight learning algorithms
  for scheduling in unknown environments,'' \emph{IEEE Transactions on
  Automatic Control}, vol.~57, no.~5, pp. 1179--1191, May. 2012.

\bibitem{zheng2009mining}
Y.~Zheng, L.~Zhang, X.~Xie, and W.-Y. Ma, ``Mining interesting locations and
  travel sequences from gps trajectories,'' in \emph{Proceedings of the 18th
  international conference on World wide web}.\hskip 1em plus 0.5em minus
  0.4em\relax ACM, 2009, pp. 791--800.

\end{thebibliography}
\end{document}